\documentclass[11pt]{article}
\usepackage{amsmath,epsfig,wrapfig}
\begin{document}
\pagestyle{empty}

\begin{flushleft}
\Large
{SAGA-HE-187-02  \hfill June 26, 2002}  \\
\end{flushleft}

\vspace{2.0cm}

\begin{center}
\LARGE
{\bf Possible Studies of} \\
\vspace{0.2cm}
{\bf Parton Distribution Functions at JHF} \\
\vspace{1.1cm}
{S. Kumano $^*$ }         \\ 
\vspace{0.4cm}
{Department of Physics}         \\
\vspace{0.1cm}
{Saga University}      \\
\vspace{0.1cm}
{Saga 840-8502, Japan} \\
\vspace{1.4cm}
\Large
{Talk given at the Workshop on} \\
\vspace{0.2cm}
{``Nuclear Physics at JHF"} \\
\vspace{0.4cm}
{KEK, Japan, May 13 -- 14, 2002} \\
{(talk on May 14, 2002)}  \\
\end{center}
\vspace{1.2cm}
\vfill
\noindent
{\rule{6.0cm}{0.1mm}} \\ 
\vspace{-0.3cm}
\normalsize
\noindent
{* Email: kumanos@cc.saga-u.ac.jp. Information on his research
       is available at}

\vspace{0.3cm}
\noindent
{\ \, \, http://hs.phys.saga-u.ac.jp.}  \\
\vspace{+0.65cm}

\hfill
{\large to be published in proceedings}
\vfill\eject
\setcounter{page}{1}
\pagestyle{plain}
\begin{center}
 
\Large

{Possible Studies of} \\

{Parton Distribution Functions at JHF} \\
 
\vspace{0.5cm}
 
{S. Kumano $^*$}             \\
 
{Department of Physics, Saga University}      \\

{Honjo-1, Saga 840-8502, Japan} \\

\vspace{0.7cm}

\normalsize
Abstract
\end{center}
\vspace{-0.60cm}

\begin{center}
\begin{minipage}[t]{10.0cm}
We discuss possible studies of parton distribution functions (PDFs) in
the nucleon and nuclei at the Japan Hadron Facility (JHF).
First, the PDFs could be investigated by the 50 GeV primary proton
facility. The distributions at medium $x$ are determined, for example,
by Drell-Yan measurements.
Second, there are feasibility studies to propose a neutrino factory
within the 50 GeV proton ring. If such an intensive high-energy neutrino
facility is built, neutrino reactions should be able to provide
valuable information on the PDFs, whereas the current structure
functions have been measured mainly for neutrino-iron reactions.

\end{minipage}
\end{center}

\vspace{0.1cm}
\section{Introduction}\label{intro}
\vspace{-0.1cm}

Structure functions have been investigated since 1970's. 
We can test both perturbative and non-perturbative aspects of Quantum
Chromodynamics (QCD) by their measurements. From measured $Q^2$ dependence
of the structure functions, we learned that it agrees with
perturbative QCD predictions. From their $x$ dependent measurements, 
parton distribution functions (PDFs) have been extracted.
They are valuable for testing non-perturbative aspects such as
predictions by lattice QCD and various phenomenological models.
Furthermore, the PDF studies have wide applications to heavy-ion physics,
neutrino oscillation experiments, and exotic event searches
at extremely large $Q^2$ beyond the current theoretical framework.

Now, the unpolarized PDFs of the nucleon are relatively well established
through a variety of experimental measurements. There are three
established groups: CTEQ (Coordinated Theoretical/Experimental Project
on QCD Phenomenology and Tests of the Standard Model),
GRV (Gl\"uck, Reya, and Vogt), and MRST (Martin, Roberts, Stirling, 
and Thorne). All the distributions agree rather well, which convinces us
that the unpolarized PDFs in the nucleon are well determined except for
extreme kinematical regions. 

On the contrary, polarized PDFs and nuclear PDFs are not reliably 
determined at this stage. The European Muon Collaboration (EMC) discovery
on the proton spin triggered enthusiastic investigations on high-energy
spin physics. Even a decade has passed since the EMC finding, we do not
understand yet how the proton spin consists of quarks and gluons.
Of course, we have rough ideas on the spin carriers mainly through the
electron and muon scattering data.
However, they are not enough to determine all the polarized PDFs.

Nuclear PDFs are in the similar situation to the polarized.
There are available experimental information from the
deep inelastic scattering (DIS) and Drell-Yan processes. They enable us
to find rough $x$ dependence of valence and antiquark distributions.
However, it is still difficult to determine gluon distributions
in nuclei. Furthermore, analysis technique has not been well
developed for the nuclear parametrization.

Considering these situations, we think that the JHF could contribute
to these PDF studies. There are two major possibilities.
One is to use the primary 50 GeV proton beam,
and the other is to use a neutrino factory. The proton beam will
certainly become available in the near future, so that proton reactions
could be used for the PDF studies. On the other hand, the possibility
of the Japanese neutrino factory is not very clear at this stage.
Neutrino factories have been considered in Europe, USA, and Japan,
so that we hope that at least one of them will be materialized. 

In this paper, we discuss possible PDF studies at JHF.
In Sec. \ref{proton}, primary proton topics are discussed. 
Then, neutrino-factory possibilities are discussed
in Sec. \ref{neutrino}. 
These topics are summarized in Sec. \ref{sum}.

\vspace{0.1cm}
\section{Primary proton beam}
\label{proton}
\vspace{-0.1cm}

The JHF is a unique facility to investigate physics associated
with secondary beams. Feasibility has been extensively studied
in such fields. On the other hand, topics associated with the
primary proton beam are not well investigated. One of the possibilities
is to use the facility for the PDF studies. Because the beam energy is 50 GeV,
we should inevitably focus on the medium and large $x$ regions of the parton
distributions. 

\vspace{0.1cm}
\subsection{Drell-Yan processes}
\label{dy}
\vspace{-0.1cm}

In the proton reactions, Drell-Yan processes provide important constraints
for the parton distributions. For example, Fermilab-E866/NuSea data 
made important contribution to the determination of $\bar u/\bar d$ asymmetry
in the nucleon. The difference between $\bar u$ and $\bar d$ was first
suggested by the New Muon Collaboration (NMC) in the studies of
Gottfried-sum-rule violation. However, the NMC data were not sufficient
to conclude actual antiquark flavor asymmetry due to the lack of
small $x$ data. The Fermilab and CERN Drell-Yan measurements indicated 
the ratio $\bar u/\bar d$ directly, and the data enabled us to determine
the difference between $\bar u$ and $\bar d$. In this way, the lepton DIS
and the Drell-Yan process are complementary in establishing precise
PDFs.

First, let us discuss the kinematical region to be probed by the possible
JHF Drell-Yan measurements. In the Drell-Yan process, an antiquark (quark) 
with momentum fraction $x_1$ in the projectile interacts with a quark (antiquark)
with $x_2$ to produce a dimuon pair with mass $m_{\mu\mu}$. 
If the center-of-mass (c.m.) energy is $\sqrt{s}$, they are related by
\begin{equation}
x_1 x_2 = \frac{m_{\mu\mu}^2}{s}
\, .
\label{eqn:x1x2}
\end{equation}
The dimuon-mass square $m_{\mu\mu}^2$ is equal to $Q^2$ of the virtual
photon, and it should be larger than a few GeV$^2$ in order to apply
perturbative QCD picture. Considering the c.m. energy $\sqrt{s}$=10 GeV,
we estimate that the target momentum fraction $x_2$ should be limited by
$x_2 > 0.1$. This means that the medium and large $x$ regions could be
investigated by the primary proton beam. 

\begin{wrapfigure}{r}{0.46\textwidth}
   \vspace{-0.0cm}
   \begin{center}
       \epsfig{file=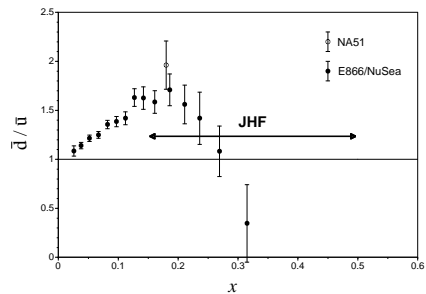,width=6.0cm}
   \end{center}
   \vspace{-0.8cm}
       \caption{\footnotesize
                Measured $\bar d/\bar u$ ratios by the E866/NuSea and NA51 
                collaborations \cite{e866,na51}.}
   \vspace{+0.3cm}
       \label{fig:dy-ud}
\end{wrapfigure}

The feasibility of such Drell-Yan measurements has been investigated
in detail by J. C. Peng {\it el al.} \cite{peng} for the JHF project.
For example, the facility could contribute to the topic of the antiquark
flavor asymmetry \cite{udbar}. The present experimental situation is shown
in Fig. \ref{fig:dy-ud}. Experimentally, the Drell-Yan cross sections for
the proton-proton and proton-deuteron reactions have been measured.
Then, the cross-section difference is converted to the antiquark flavor
asymmetry in Fig. \ref{fig:dy-ud}. The experimental data are taken by
the E866/NuSea \cite{e866} and NA51 \cite{na51} collaborations. 
It is clear from the figure that $\bar d$ is larger than $\bar u$
in the $x$ region, $x=0.04 \sim 0.2$. Possible JHF experiments could
extend the measured region to larger $x$, where current data are not
accurate enough or even not available.

\begin{figure}[h]
\parbox[t]{0.46\textwidth}{
   \vspace{+0.1cm}
   \begin{center}
       \epsfig{file=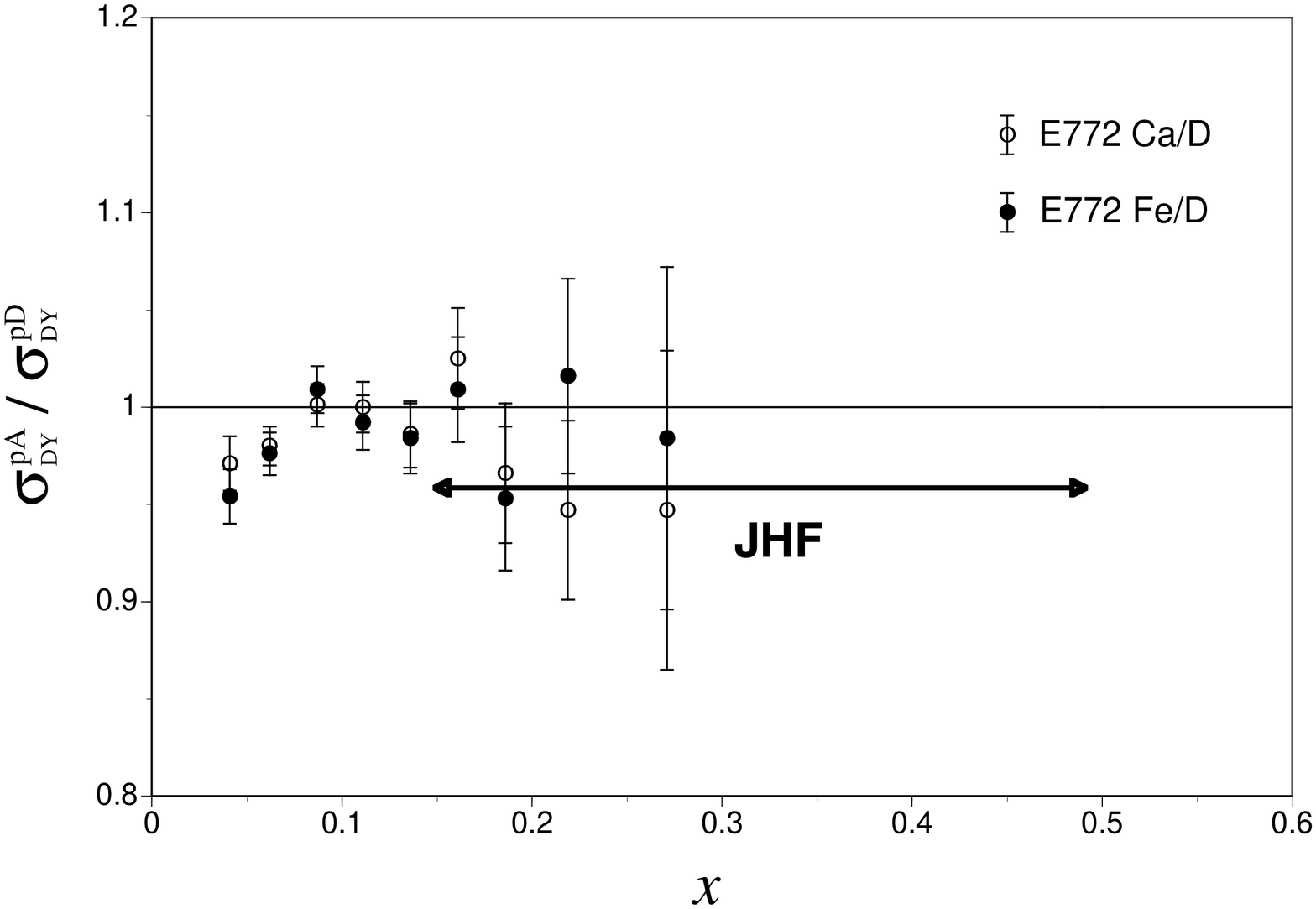,width=6.0cm}
   \end{center}
   \vspace{-0.8cm}
       \caption{\footnotesize
                Nuclear Drell-Yan cross section ratios \cite{e772}.}
       \label{fig:dy-nucl}
}\hfill
\parbox[t]{0.46\textwidth}{
   \vspace{-0.0cm}
   \begin{center}
       \epsfig{file=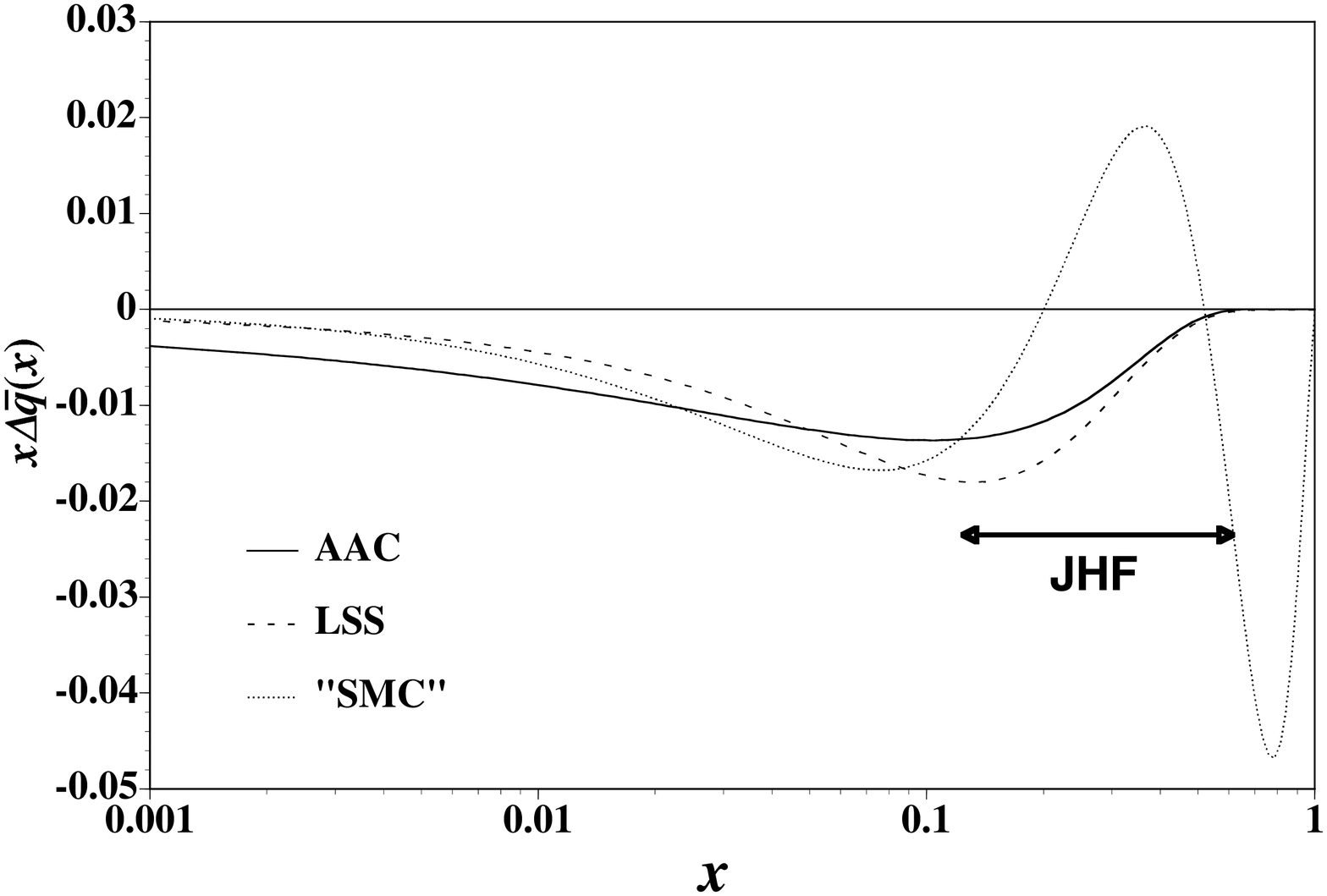,width=6.0cm}
   \end{center}
   \vspace{-0.8cm}
       \caption{\footnotesize
                Antiquark distributions by LSS, SMC, and AAC \cite{aac1}.}
       \label{fig:dy-pol}
}
\end{figure}

Next, the Drell-Yan experiments could be used for determining nuclear
antiquark distributions in the medium $x$ region. 
The situation of determining nuclear PDFs is much worse than the one for
the nucleon in the sense that a variety of experimental data are not available
and that analysis technique has not been well developed.
For example, although the nuclear antiquark distributions could be roughly
determined by the measurements of the $F_2$ structure functions at small $x$,
their determination is not possible at medium $x$ due to the lack of data.
The present situation is typically shown in Fig. \ref{fig:dy-nucl},
where E772 data are shown for the cross-section ratios,
$\sigma^{pCa}/\sigma^{pD}$ and $\sigma^{pFe}/\sigma^{pD}$ \cite{e772}.
We should note that the cross-section
ratios are almost equal to the antiquark ratios, 
$\bar q ^{\, Ca}/\bar q^{\, D}$ and $\bar q ^{\, Fe}/\bar q^{\, D}$
in the region, $x<0.1$. The data indicate small nuclear corrections 
in the antiquark distributions at $x \sim 0.1$.
However, they are not accurate enough at larger $x$, so that
the determination is almost impossible in the medium $x$ region.
This fact makes a nuclear $\chi^2$ analysis difficult for
fitting various experimental data \cite{sagapdf}.
In the parametrization of the nuclear PDFs \cite{sagapdf,ekr},
nuclear $F_2$ data are used together with the Drell-Yan data to produce
the optimum distributions. If the medium $x$ data are taken at JHF
as indicated in Fig. \ref{fig:dy-nucl}, the antiquark distributions are
easily constrained at medium $x$, and the $\chi^2$ analysis becomes
much reliable.

\vspace{-0.5cm}
\noindent

Another application is for polarized PDFs. The polarized distributions in
the nucleon have been obtained by using polarized electron and muon DIS
experimental data. They are taken mainly for the structure function $g_1$,
but there are also available semi-inclusive data. 
Several parametrizations have been proposed for explaining the polarized
experimental data. However, the data are still taken in a limited kinematical
range, and a variety of data are not available yet. 
This situation makes accurate determination of polarized
antiquark distributions difficult. It is clearly illustrated in
Fig. \ref{fig:dy-pol} \cite{aac1}. Three antiquark distributions are shown
in the figure, and they are obtained by Leader-Sidrov-Stamenov (LSS),
Spin Muon Collaboration (SMC), and Asymmetry Analysis Collaboration (AAC).
It is obvious from the figure that the antiquark distributions at medium $x$
are much different depending on the analysis methods.
The JHF project probes the medium $x$ range as shown in Fig. \ref{fig:dy-pol}
if the proton is polarized, so that its data are also valuable for 
the high-energy spin physics, especially for the determination of
$\Delta \bar q$ at medium $x$.

\vspace{0.1cm}
\subsection{Drell-Yan processes with polarized deuteron target}
\label{dy-pd}
\vspace{-0.1cm}

Using a polarized deuteron target, we could investigate polarized
proton-deuteron processes. The reactions have never been investigated
experimentally, whereas there are a few theoretical studies.
There are two major purposes. One is to investigate
new tensor structure functions, which do not exist for the spin-1/2
nucleon, and the other is to find flavor asymmetry for the polarized
antiquark distributions. 

First, we discuss the tensor structure functions. 
The polarized proton-proton ({\it pp}) reactions have been extensively
investigated theoretically, and experimental studies are in progress
as the Relativistic Heavy Ion Collider (RHIC) Spin project.
We expect that nucleon spin structure will be partially clarified
in the near future. However, it is important that different aspects
of spin physics should be also investigated in order to test our knowledge
of high-energy spin. One of such quantities is the tensor structure
of the deuteron, and it could be studied by the tensor structure
functions \cite{b1}.
In the electron and muon scattering, there are leading-twist structure
functions $b_1$ and $b_2$, which are related by the Callan-Gross type
relation $2x b_1=b_2$. 
The antiquark part of the tensor structure could be investigated
by polarized proton-deuteron ({\it pd}) Drell-Yan processes
in the same way as the unpolarized Drell-Yan reactions
in Sec. \ref{dy}.

A theoretical formalism of the polarized $pd$ Drell-Yan was investigated
in Ref. \cite{pd-dy}. In comparison with the $pp$ Drell-Yan processes,
there are additional structure functions associated with the tensor structure
of the deuteron. Among them, a quadrupole spin asymmetry $A_{UQ_0}$ is sensitive
to the $b_1$ distributions:
\begin{equation}
A_{UQ_0}  =  \frac{\sum_a e_a^2 \, 
                  \left[ \, f_1(x_1) \, \bar b_1(x_2)
                          + \bar f_1(x_1) \, b_1(x_2) \, \right] }
                {\sum_a e_a^2 \, 
                  \left[ \, f_1(x_1) \, \bar f_1(x_2)
                          + \bar f_1(x_1) \, f_1(x_2) \, \right] }
\, .
\label{eqn:auq0}
\end{equation}
The notation $UQ_0$ indicates that the proton is unpolarized and
a tensor spin combination is taken for the deuteron.
The functions $f_1(x)$ and $\bar f_1(x)$ are unpolarized quark and
antiquark distributions, and $b_1(x)$ and $\bar b_1(x)$ 
are tensor-polarized distributions. In the large $x_F$ region,
the spin asymmetry becomes
\begin{equation}
A_{UQ_0} \textrm{(large $x_F$)} 
      \approx \frac{\sum_a e_a^2 \, f_1(x_1) \, \bar b_1(x_2)}
                   {\sum_a e_a^2 \, f_1(x_1) \, \bar f_1(x_2)}
\ \ \ \text{at large $x_F$} \, .
\end{equation}
It indicates that the tensor polarized antiquark distributions can be
obtained by the spin asymmetry measurements in the polarized $pd$
reactions. This is complementary to the studies of the tensor structure
in the electron and muon scattering, where the antiquark distributions
cannot be determined in the medium $x$ region. 

Second, the polarized $pd$ Drell-Yan could be used for investigating
the polarized antiquark flavor asymmetry $\Delta \bar u/\Delta \bar d$
\cite{km}. As shown in Sec. \ref{dy}, the difference between the $pp$
and $pd$ cross sections is attributed to the difference between $\bar u$
and $\bar d$. In the same way, the difference between the polarized
$pp$ and $pd$ cross sections should be associated with the difference
between $\Delta \bar u$ and $\Delta \bar d$. Another important point
is that it enables us to determine $\Delta_T \bar u/\Delta_T \bar d$
for the transversity distributions, because the transversity cannot
be measured by the $W$ production process.
The cross-section ratio for the $pp$ and $pd$ reactions is 
\small
\begin{equation}
R_{pd} \equiv \frac{     \Delta_{(T)} \sigma_{pd}}
                   {2 \, \Delta_{(T)} \sigma_{pp}}
        =     \frac{ \sum_a e_a^2 \, 
    \left[ \, \Delta_{(T)} q_a(x_1) \, 
              \Delta_{(T)} \bar q_a^{\, d}(x_2)
            + \Delta_{(T)} \bar q_a(x_1) \, 
              \Delta_{(T)} q_a^d(x_2) \, \right] }
              { 2 \, \sum_a e_a^2 \, 
    \left[ \, \Delta_{(T)} q_a(x_1) \, 
              \Delta_{(T)} \bar q_a(x_2)
            + \Delta_{(T)} \bar q_a(x_1) \, 
              \Delta_{(T)} q_a(x_2) \, \right] }
\, ,
\label{eqn:ratio1}
\end{equation}
where $\Delta_{(T)} q$ denotes a polarized distribution
$\Delta q$ or $\Delta_T q$ depending on
the longitudinal or transverse polarization.
If the large $x_F$ ($=x_1-x_2$) region is considered,
the ratio becomes
\small
\begin{equation}
R_{pd} (x_F\rightarrow 1)    =  \frac{1}{2} \, \left [ \, 1 
                 + \frac{\Delta_{(T)} \bar d (x_2)}
                        {\Delta_{(T)} \bar u (x_2)} 
                    \, \right ]_{x_2\rightarrow 0}
\, ,
\label{eqn:rpd+1}
\end{equation}
\normalsize
which is directly proportional to the flavor asymmetry ratio.
It indicates that the flavor asymmetry could be measured especially
at large $x_F$.

\vspace{0.1cm}
\subsection{Comments on other processes}
\label{other}
\vspace{-0.1cm}

We have discussed the PDF studies by the Drell-Yan processes; however,
there are other possibilities. For example, $J/\psi$ production
could be used for investigating the antiquark flavor asymmetry \cite{peng}.
Furthermore, the process is known to be sensitive to the gluon distribution,
so that the production measurements could impose constraints on the 
unpolarized and polarized gluon distributions.

On the other hand, the feasibility of direct photon measurements has
not been studied for the JHF project. Since the beam energy is not high,
direct photons from perturbative processes could be mixed with other
background contributions. However, if the measurements are possible,
the JHF contributes to the gluon determination at medium $x$.
Several years ago, the CDF (Collider Detector at Fermilab) group
reported anomalous events in their jet cross sections for
the $p+\bar p$ reaction at $\sqrt{s}=1.8$ TeV.
Because they cannot be explained by the next-to-leading-order (NLO) QCD,
the events were considered to be a signature beyond the current
theoretical framework \cite{cdf}. However, it was revealed that
the gluon distribution at large $x$ could be the reason for the discrepancy
from the QCD prediction. In fact, according to Ref. \cite{cteq4},
the events could be explained by a gluon enhancement at large $x$.
We think that unexpected events for new physics should be found
at extremely large $Q^2$, which inevitably means large $x$.
As shown by the anomalous CDF events and also similar ones at HERA,
the PDFs at large $x$ should be known precisely in order to
find any exotic signatures. In this sense, the JHF has potential to
contribute to such a large $x$ determination.

\vspace{0.1cm}
\section{Neutrino factory}
\label{neutrino}
\vspace{-0.1cm}

In addition to the primary proton beam, there is another passibility
for the PDF studies at JHF by using neutrinos from accelerated muons.
It is called a neutrino factory. Its feasibility studies are still 
at the early stage. However, we hope that at least one of the neutrino
factory plans in Europe, USA, and Japan will be realized.
At this stage, a 30 GeV neutrino factory is considered
as a possibility within the 50 GeV proton ring.
In neutrino-nucleon scattering, the Bjorken scaling variable is
given by $x=Q^2/(2 M_N \nu)$, where $M_N$
is the nucleon mass and $\nu$ is the energy transfer. 
In order to be considered as DIS data, $Q^2$ should be larger than
at least 1 GeV$^2$, and $\nu$ should be smaller than the neutrino
beam energy. Therefore, the minium $x$ is 
$x_{min} = 1 / (2 \cdot 1 \cdot 30) \approx 0.017$.
In the following, we discuss nucleon and nuclear PDF studies
at such a neutrino facility.

\vspace{0.1cm}
\subsection{Comments on unpolarized structure functions in the nucleon}
\label{n}
\vspace{-0.1cm}

There have been already measurements of neutrino DIS. In fact, the data
have been included in the unpolarized PDF parametrizations. Especially,
the opposite sign dimuon events are crucial for the determination of the
strange quark distribution, and $F_3$ structure functions are valuable
for the valence-quark distributions. In the recent parametrization,
the CCFR (Columbia-Chicago-Fermilab-Rochester) data have been used. 
It should be, however, noticed that the target is iron instead of the proton. 
Although some nuclear corrections, which are suggested by nuclear $F_2$
modification, are applied, it is not obvious that proper PDFs
in the ``nucleon" have been extracted from the CCFR iron data. 
A future neutrino factory could clarify the distributions in the nucleon
if the proton and deuteron targets could be used with an intense neutrino
beam. 

\vspace{0.1cm}
\subsection{Polarized structure functions}
\label{pol}
\vspace{-0.1cm}

Polarized structure functions $g_1$ and $g_2$ have been measured
by electron and muon DIS. In the neutrino scattering,
there are additional ones, $g_3$, $g_4$, and $g_5$,
due to a parity-violation nature of the reaction. 
These additional structure functions have not been measured 
experimentally, and there are not sufficient theoretical studies.
It is typically reflected in confusing definitions of 
$g_3$, $g_4$, and $g_5$. Various definitions are summarized in
Ref. \cite{g345}, where we notice that someone's $g_3$ structure function 
could be $g_5$ for some others and vice versa.
In reading the papers on these structure functions, one should be
careful about their definitions. In the following, the
conventions of Ref. \cite{lr} are used. In the leading order (LO),
they are related to the polarized parton distributions as
\begin{align}
(g_4^{\nu p}+g_5^{\nu p})/2x & = g_3^{\nu p}
 = - (\Delta d + \Delta s - \Delta \bar u - \Delta \bar c) ,
\nonumber \\ 
(g_4^{\bar\nu p}+g_5^{\bar\nu p})/2x & = g_3^{\bar \nu p} 
 = -(\Delta u + \Delta c - \Delta \bar d - \Delta \bar s) .
\end{align}
Combining these structure functions, we have
\begin{align}
g_3^{\nu p} + g_3^{\bar \nu p} = &
-  ( \Delta u_v + \Delta d_v ) - (\Delta s -\Delta \bar s)
-  (\Delta c -\Delta \bar c) ,
\nonumber \\ 
\frac{1}{2} \, [ \, g_3^{\bar \nu (p+n)} - g_3^{\nu (p+n)} \, ] = &
(\Delta s +\Delta \bar s) - (\Delta c +\Delta \bar c) .
\label{eqn:g3}
\end{align}
The differences $\Delta s -\Delta \bar s$ and $\Delta c -\Delta \bar c$
are expected to be small, so that the first combination 
$g_3^{\nu p} + g_3^{\bar \nu p}$ could be used for determining 
the polarized valence-quark distributions. The second combination
$g_3^{\bar \nu (p+n)} - g_3^{\nu (p+n)}$ is sensitive to the strange
and charm distributions.

There is additional importance in the neutrino scattering.
The quark spin content of the nucleon has been controversial for a decade.
The EMC finding initiated many theoretical and experimental investigations.
Despite such efforts, the spin content $\Delta \Sigma$ is not still clear.
As shown in Fig. \ref{fig:dy-pol}, the obtained antiquark distributions
have different $x$ dependence in the small $x$ region due to the lack of
small $x$ data. It makes the determination of $\Delta \Sigma$ rather
difficult. In fact, the obtained $\Delta \Sigma$ ranges from 5\%
to 30\% depending on the models in Fig. \ref{fig:dy-pol}. 
In order to clarify $\Delta \Sigma$, the neutrino reactions are very
useful. The $g_1$ structure functions are expressed in the parton model as
\begin{align}
g_1^{\nu p} = & \Delta d + \Delta s + \Delta \bar u + \Delta \bar c ,
\nonumber \\ 
g_1^{\bar \nu p} = & \Delta u + \Delta c + \Delta \bar d + \Delta \bar s .
\end{align}
Combining these expressions, we find
\begin{equation}
\int dx \, (g_1^{\nu p}+g_1^{\bar \nu p}) = \Delta \Sigma ,
\end{equation}
in the LO. Although there are NLO and higher-order corrections,
the spin content could be obtained directly without resorting to
low-energy semi-leptonic data for fixing the first moments of
the valence-quark distributions.

In this way, if the neutrino factory is realized in future at JHF,
it could contribute to the high-energy spin physics significantly.

\vspace{0.1cm}
\subsection{Nuclear structure functions}
\label{nucl}
\vspace{-0.1cm}

We mentioned that many of the actual neutrino data are taken
for the iron target, so that neutrino-nucleus DIS data already
exist. However, experimental nuclear modification has not been discussed
seriously for the neutrino reactions because accurate deuteron data
are not obtained yet. At the neutrino factory, accurate deuteron
data are expected. It is especially interesting to investigate
nuclear $F_3$ structure functions which are specific for parity-violating
neutrino scattering. For example, combining functions of different
reactions, we obtain
\begin{equation}
\frac{1}{4} \, [ \, F_3^{\nu (p+n)} + F_3^{\bar \nu (p+n)} \, ]
= \, u_v+d_v + (s - \bar s) + (c - \bar c) 
\approx \, u_v+d_v 
\ ,
\end{equation}
for investigating the valence-quark distributions.

\begin{wrapfigure}{r}{0.46\textwidth}
   \vspace{-0.0cm}
   \begin{center}
       \epsfig{file=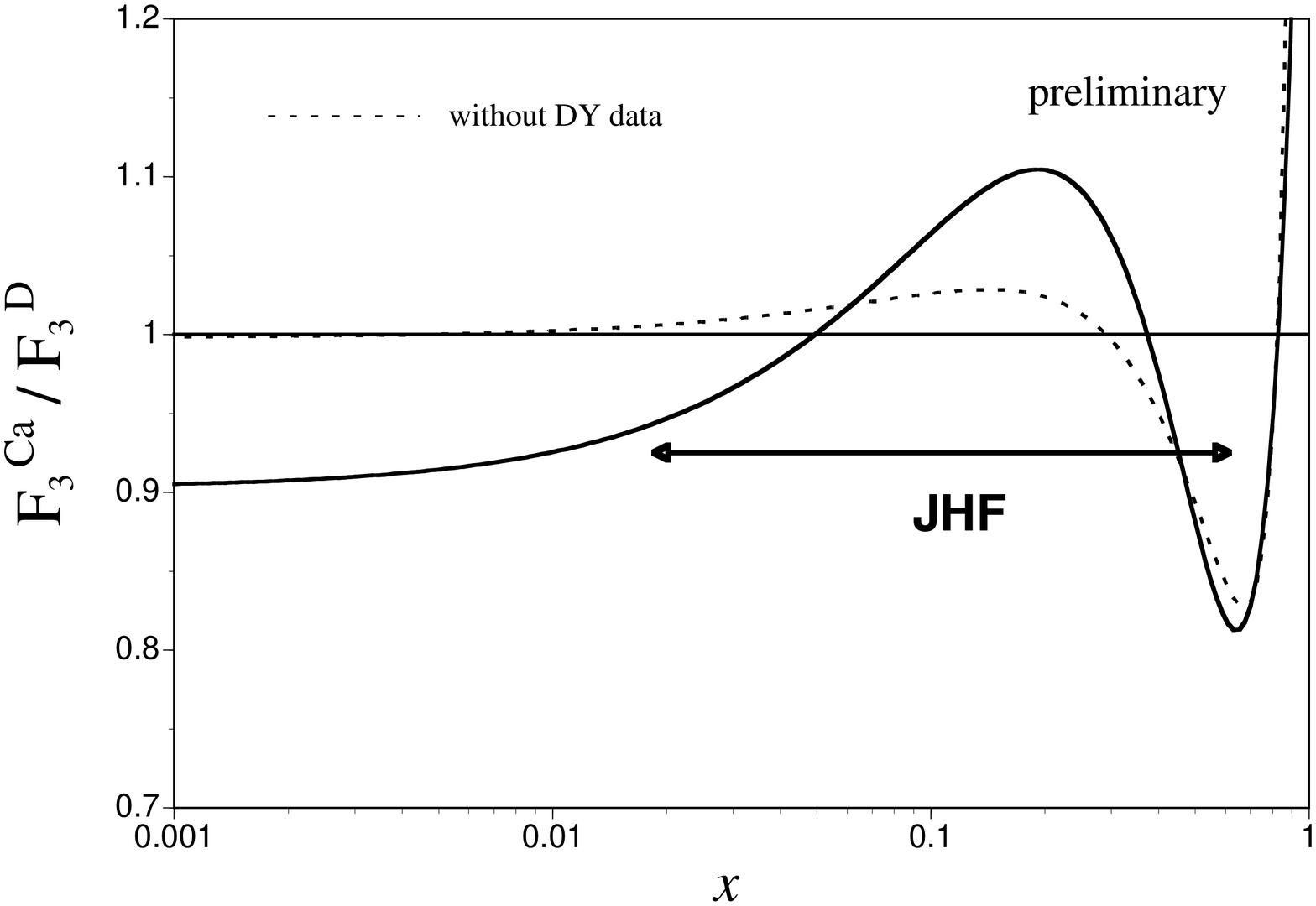,width=6.0cm}
   \end{center}
   \vspace{-0.7cm}
       \caption{\footnotesize
         Nuclear modification of the $F_3$ structure function \cite{sagapdf}.}
       \label{fig:f3}
\end{wrapfigure}

The nuclear PDFs can be determined by analyzing nuclear DIS data 
together with the Drell-Yan data \cite{sagapdf,ekr}. The obtained valence-quark
distributions are shown in Fig. \ref{fig:f3}, where the curves indicate
$q_v^{Ca}/q_v^{D}$. The dashed curve is obtained by removing the Drell-Yan data
from the analysis data set. The neutrino factory should provide strong
constraints on the $x$ dependence in the region $x > 0.02$. 
In addition, if the beam energy is large enough, the shadowing region
could be probed. In the $F_2$ structure functions for the electron
and muon DIS, the shadowing phenomenon is well known. However,
there is no experimental information on the $F_3$ shadowing.
Although the valence-quark distributions are constrained by the baryon number
and charge, there is no strong restriction on the small $x$ behavior.
In principle, anti-shadowing is not completely ruled out
in the present situation. Experimental measurements of the $F_3$ shadowing
or anti-shadowing phenomenon are important for understanding high-energy
nuclear reactions precisely \cite{f3}.

\vspace{0.1cm}
\section{Summary}
\label{sum}
\vspace{-0.1cm}

We have explained possible measurements of structure functions and
parton distribution functions in the nucleon and nuclei at JHF.
There are two possibilities. On is to use the primary proton beam,
and the other is to use the neutrino factory. These facilities should
provide important data, which cannot be obtained by other laboratories.
The Drell-Yan measurements should clarify the antiquark distributions
at medium $x$, especially the flavor asymmetry ratios $\bar u/\bar d$, 
$\Delta \bar u/\Delta \bar d$, and $\Delta_T \bar u/\Delta_T \bar d$.
Furthermore, nuclear antiquark distributions should be determined
in the medium $x$ region. We hope that feasibility 
of other processes, such as the charmonium and direct-photon productions, 
will be seriously studied in the near future.

The Japanese neutrino factory has been considered at JHF.
Although long baseline physics (neutrino oscillation)
has been rather well studied in the Japanese particle physics community,
short baseline physics (hadron structure) has not been well discussed.
The facility could be important for finding the details of the polarized PDFs
especially in connection with the quark spin content. There are also
new structure functions $g_3$, $g_4$, and $g_5$ which do not exit
in the electron and muon scattering. Furthermore, the nuclear valence-quark
distributions should be clarified by the $F_3$ structure functions in nuclei.
It is particularly interesting to find whether the $F_3$ shadowing is
similar to the observed $F_2$ shadowing.
 
\vspace{0.5cm}

\noindent
{* Email: kumanos@cc.saga-u.ac.jp. URL: http://hs.phys.saga-u.ac.jp.} \\

\vspace{-0.5cm}


\begin{thebibliography}{99}
\vspace{-0.2cm}
\bibitem{peng}   J.-C. Peng, G. T. Garvey, J. M. Moss, S. Sawada, 
                    and J. Chiba, hep-ph/0007341.
\vspace{-0.2cm}
\bibitem{udbar} S. Kumano, Phys. Rep. 303 (1998) 183;
                J.-C. Peng and G. T. Garvey,
                      in {\it Trend in Particle and Nuclear Physics}, Volume 1, 
                       Plenum Press. 
\vspace{-0.2cm}
\bibitem{e866}  E866/NuSea collaboration,$\,$R.S.$\,$Towell {\it et al.},
                    Phys.Rev.D64 (2001) 052002. 
\vspace{-0.2cm}
\bibitem{na51}  NA51 collaboration, A. Baldit {\it et al.}, 
                    Phys. Lett. B332 (1994) 244.
\vspace{-0.2cm}
\bibitem{e772}  E772 collaboration, D. M. Alde {\it et al.},
                      Phys. Rev. Lett. {\bf 64} (1990) 2479.
\vspace{-0.2cm}
\bibitem{sagapdf} M. Hirai, S. Kumano, and M. Miyama,
                     Phys. Rev. D64 (2001) 034003;
                  M. Hirai and S. Kumano, research in progress.
                  See http://hs.phys.saga-u.ac.jp/nuclp.html.
\vspace{-0.2cm}
\bibitem{ekr} K. J. Eskola, V. J. Kolhinen, and P. V. Ruuskanen,
                     Nucl. Phys. B535 (1998) 351;
               K. J. Eskola {\it et al.}, hep-ph/0110348.
\vspace{-0.2cm}
\bibitem{aac1}  Asymmetry Analysis Collaboration (AAC), 
                Y. Goto {\it et al.}, Phys. Rev. D62 (2000) 034017.
                See http://spin.riken.bnl.gov/aac/.
\vspace{-0.2cm}
\bibitem{b1}   P. Hoodbhoy, R. L. Jaffe, and A. Manohar, 
                       Nucl. Phys. B312 (1989) 571;
               F. E. Close and S. Kumano, Phys. Rev. D42 (1990) 2377.
\vspace{-0.2cm}
\bibitem{pd-dy} S. Hino and S. Kumano, Phys.Rev. D59 (1999) 094026; 
                                                  D60 (1999) 054018.
\vspace{-0.2cm}
\bibitem{km}    S. Kumano and M. Miyama, Phys. Lett. B479 (2000) 149.
\vspace{-0.2cm}
\bibitem{cdf}   CDF collaboration,  F. Abe {\it et al.},
                          Phys. Rev. Lett. 77 (1996) 438.
\vspace{-0.2cm}
\bibitem{cteq4} CTEQ collaboration,  H. L. Lai {\it et al.},
                          Phys. Rev. D55 (1997) 1280.
\vspace{-0.2cm}
\bibitem{g345}  J. Bl\"umlein and M. Kochelev, Nucl. Phys. B498 (1997) 285.
\vspace{-0.2cm}
\bibitem{lr}    B. Lampe and E. Reya, Phys. Rep. 332 (2000) 1.
\vspace{-0.2cm}
\bibitem{f3}  L. L. Frankfurt, M. I. Strikman, and S. Liuti,
                       Phys. Rev. Lett. 65 (1990) 1725;
              R. Kobayashi, S. Kumano, and M. Miyama, 
                       Phys. Lett. B354 (1995) 465;
              S. A. Kulagin, Nucl. Phys. A640 (1998) 435.
\end{thebibliography}
\end{document}